\documentclass[fleqn,usenatbib]{mnras}

\usepackage[T1]{fontenc}

\DeclareRobustCommand{\VAN}[3]{#2}
\let\VANthebibliography\thebibliography
\def\thebibliography{\DeclareRobustCommand{\VAN}[3]{##3}\VANthebibliography}

\usepackage{graphicx}	
\usepackage{amsmath}	
\usepackage{amssymb}	

\usepackage{lipsum}

\newcommand{\gasfrac}{$f_{\mathrm{CGM}}$}

\newcommand{\msol}{{M$_\odot$}}

\newcommand{\kappacogas}{$\kappa_{\rm co,gas}$}
\newcommand{\kappacostar}{$\kappa_{\rm co,\star}$}

\newcommand{\organic}{{\sc organic}}
\newcommand{\enhanced}{{\sc enhanced}}
\newcommand{\suppressed}{{\sc suppressed}}

\usepackage{newtxtext,newtxmath}

\title[Galaxy mergers, the baryon cycle and quenching]{Galaxy mergers can initiate quenching by unlocking an AGN-driven transformation of the baryon cycle}

\author[J. J. Davies, A. Pontzen and R. A. Crain]{Jonathan  J. Davies,$^{1}$\thanks{E-mail: j.j.davies@ucl.ac.uk}
Andrew Pontzen$^{1}$
and Robert A. Crain,$^{2}$
\\
$^{1}$Department of Physics and Astronomy, University College London, Gower Street, London WC1E 6BT, UK\\
$^{2}$Astrophysics Research Institute, Liverpool John Moores University, 146 Brownlow Hill, Liverpool L3 5RF, UK
}

\date{Accepted XXX. Received YYY; in original form ZZZ}

\pubyear{2022}

\begin{document}
\label{firstpage}
\pagerange{\pageref{firstpage}--\pageref{lastpage}}
\maketitle
\begin{abstract}
We use zoom simulations to show how merger-driven disruption of the gas disc in a galaxy provides its central active galactic nucleus (AGN) with fuel to drive outflows that entrain and expel a significant fraction of the circumgalactic medium (CGM). This in turn suppresses replenishment of the interstellar medium, causing the galaxy to quench up to several Gyr after the merger. We start by performing a zoom simulation of a present-day star-forming disc galaxy with the EAGLE galaxy formation model. Then, we re-simulate the galaxy with controlled changes to its initial conditions, using the genetic modification technique. These modifications either increase or decrease the stellar mass ratio of the galaxy's last significant merger, which occurs at $z\approx 0.74$. The halo reaches the same present-day mass in all cases, but changing the mass ratio of the merger yields markedly different galaxy and CGM properties. We find that a merger can unlock rapid growth of the central supermassive black hole if it disrupts the co-rotational motion of gas in the black hole's vicinity. Conversely, if a less disruptive merger occurs and gas close to the black hole is not disturbed, the AGN does not strongly affect the CGM, and consequently the galaxy continues to form stars. Our result illustrates how a unified view of AGN feedback, the baryon cycle and the interstellar medium is required to understand how mergers and quenching are connected over long timescales.
\end{abstract}




\section{Introduction}

A connection between rapid black hole (BH) growth and galaxy mergers is a long-standing theoretical prediction of galaxy formation models, with evidence provided by idealised numerical simulations of merging galaxy pairs \citep[e.g.][]{dimatteospringelhernquist05,springeldimatteohernquist05,hopkins06,younger09,hopkins10a}, analytical and semi-analytical modelling \citep[e.g.][]{draperballantyne12,menci14}, and zoom-in and large-volume cosmological simulations \citep[e.g.][]{sijacki07,sijacki15,bellovary13,dubois15,steinborn18,mcalpine20}. Conclusive observational evidence of such a connection remains elusive, however, as studies that differ in their selection of active galactic nucleus (AGN) hosts and merging galaxies yield conflicting results, from a strong merger-AGN connection to no connection at all \citep[see e.g.][and references therein]{ellison19}.

The AGN feedback associated with merger-induced BH growth could have a transformative effect on the star formation activity in the remnant galaxy. AGN are observed to drive mass-loaded outflows that could significantly deplete their host galaxies of interstellar gas \citep[e.g.][]{rupke11,maiolino12,harrison14,fluetsch19,forster19,circosta21,concas22}, and the BH mass (or a proxy for it) is the best predictor of quenching in galaxy surveys \citep[e.g.][]{kauffmann06,cheung12,fang13,terrazas16,terrazas17,bluck20,piotrowska21}. While AGN feedback has long been recognised as the most likely means of regulating the growth of massive galaxies \citep[e.g.][]{bower06,croton06,mccarthy10,fabian12}, cosmological simulations such as EAGLE, IllustrisTNG and SIMBA have only recently revealed an intimate connection between BH growth and the star formation rates of lower mass, Milky Way-like galaxies, with the impact of AGN feedback on the circumgalactic medium playing a key role in this connection.

In these simulations, correlations exist (at fixed halo mass) between the central BH mass, halo gas content and galaxy star formation rate \citep{davies19,davies20,dave19,terrazas20,appleby21,cui21,robsondave21}, and the cooling time of the CGM \citep{davies20,zinger20}, such that quenched galaxies typically host overmassive BHs and have gas-poor and inefficiently-cooling circumgalactic reservoirs. Together, these correlations suggest that for quiescence to be achieved and maintained in Milky Way-like galaxies, AGN feedback must eject a large fraction of the CGM over the galaxy's lifetime, preventing the CGM from efficiently cooling and replenishing the galaxy's gas supply for star formation. This connection remains challenging to demonstrate through observations, though analysis of soft X-ray maps from the eFEDS survey provides encouraging evidence that star-forming galaxies with stellar masses similar to the Milky Way have brighter X-ray atmospheres than quenched galaxies of similar mass, suggesting that their haloes are more gas-rich \citep{chadayammuri22}.

Disruptive events such as galaxy mergers could play a key role in driving the BH growth and AGN feedback needed to produce these correlations. In addition to the aforementioned theoretical studies linking mergers and rapid BH growth, this hypothesis is supported by other, related correlations in the EAGLE and IllustrisTNG galaxy populations; at fixed halo mass, galaxies hosted by gas-rich haloes tend to be star-forming discs, while those with gas-poor haloes tend to exhibit disturbed, dispersion-dominated kinematics \citep{davies20}, indicative of an earlier merger or disruption event. Through the AGN feedback they induce, galaxy mergers could have a transformative effect on the content and physical state of the CGM, fundamentally altering the baryon cycle of the galaxy-halo ecosystem. In this study, we investigate how the CGM of a Milky Way-like galaxy is affected by a merger event, and examine the consequences for the evolution of the remnant galaxy.

We do so by performing a controlled experiment on a single galaxy, simulated in its full cosmological context using the `zoom' technique. By using the genetic modification (GM) technique \citep{roth16}, we are able to systematically adjust the stellar mass ratio of an individual merger that occurs in this galaxy's lifetime, causing a more disruptive event or even preventing it from happening altogether, whilst preserving the overall halo mass accretion history and environment of the galaxy \citep[see e.g.][]{pontzen17,sanchez19,sanchez21}. 

By using this method to probe how mergers impact the baryon cycle, we disentangle different effects in a way that is not possible with statistical studies. The connection between mergers, AGN and the CGM may depend on the halo and stellar masses of the participating galaxies, the luminosity of the AGN, the identification of merging systems, the properties of the environment, and the halo accretion and merger histories. By examining the evolution of a single galaxy and adjusting one variable (in this case, the stellar mass ratio of a merger), we can keep the environment fixed and reveal in detail how each of the aforementioned quantities changes in response to this adjustment. This enables us to establish a causal link between this merger, AGN feedback, and the evolution of the galaxy and its CGM.

We begin by describing our numerical experiment, the simulation code employed, and our analysis techniques in Section \ref{sec:methods}. We evaluate the mass assembly and merger histories of our galaxy and its genetically-modified counterparts in Section \ref{sec:results:mass}, before investigating how a merger affects the evolution of the central BH and the CGM in Section \ref{sec:results:bhCGM}. We examine how mergers can drive high BH accretion rates in Section \ref{sec:results:vicinity}, and explore how the associated AGN feedback impacts the thermodynamic properties of the CGM in Section \ref{sec:results:preventative}. Finally, we explore the consequences of this process for the evolution of the galaxy in Section \ref{sec:results:galaxy}, and summarise and discuss our results in Section \ref{sec:summary}.

\section{Methods}
\label{sec:methods}

In this study, we produce a suite of simulations evolved using the \textsc{EAGLE} version of the gravity and smoothed-particle hydrodynamics code \textsc{gadget3} \citep[last described by][]{springel05}. We employ the `zoom' simulation technique \citep[e.g.][]{katz93,bertschinger01} to follow the evolution of an individual galaxy and its immediate environment at high resolution, while also simulating a large-scale periodic volume using collisionless gravitational dynamics at low resolution, thus preserving the correct large-scale forces. In this section, we describe how our candidate galaxy was selected and how its zoom initial conditions were constructed and evolved (Section \ref{sec:methods:ics}), outline how its genetically-modified variants were designed (Section \ref{sec:methods:gm}) and describe how we identify and track galaxies and haloes within the simulations (Section \ref{sec:methods:tangos}).

\subsection{Construction and evolution of the initial conditions}
\label{sec:methods:ics}

In this study, our unmodified fiducial galaxy (henceforth referred to as the \organic{} system) is the same system that was studied by \citet{davies21}, to which we refer the reader for further details. In brief, we selected a present-day central disc galaxy with stellar mass $M_\star=4.3\times 10^{10}$ M$_\odot$ that is hosted by a halo of mass $M_{200}=3.4\times 10^{12}$ \msol{}. We focus on this halo mass scale as this is where correlations between the central BH masses, halo gas contents and specific star formation rates of present-day EAGLE central galaxies are the strongest \citep{davies20}. We selected this particular galaxy because it lies on the star-forming main sequence (sSFR$=10^{-10.2}$ yr$^{-1}$) and has a CGM mass fraction, $f_{\rm CGM}$ (normalised to the cosmic baryon fraction $\Omega_{\rm b}/\Omega_0$), of 0.31, which is close to the median value at this mass in the flagship EAGLE volume \citep{davies20}. We define $f_{\rm CGM}\equiv M_{\rm CGM}/M_{200}$, where $M_{\rm CGM}$ is the total mass of all gas within the virial radius that is not star-forming.

The galaxy was not selected from the publicly-available suite of EAGLE simulations, but from a separate periodic simulation volume, which was evolved using the Reference EAGLE simulation model from initial conditions generated with the \textsc{genetIC} software \citep{stopyra20}. This parent volume adopts the cosmological parameters from \citet{planck16}, is $L=50$ cMpc on a side, and contains $512^3$ dark matter particles of mass $3.19\times 10^7$ \msol{}  and, initially, an equal number of baryonic particles of mass $5.6\times 10^6$ \msol{}, a particle resolution similar to that of the standard EAGLE simulation suite.

To generate our zoomed initial conditions, we select all particles within a sphere of radius $r=3r_{200}$ centred on our candidate galaxy, and identify the Lagrangian region defined by these particles in the initial conditions. We then refine this Lagrangian region using a factor of 8 more particles, while also downsampling the remainder of the simulation volume by a factor of 8, yielding particle masses of $m_{\rm g} = 7.35 \times 10^{5}$ \msol{}, $m_{\rm dm} = 3.98 \times 10^{6}$ \msol{}, and $m_{\rm lr} = 3.02 \times 10^{8}$ \msol{} for gas, dark matter and low-resolution mass tracer particles respectively.

We evolve these initial conditions with the EAGLE model \citep{schaye15,crain15,mcalpine16}. We adopt the recalibrated (Recal) parameter values defined by \citet{schaye15} for EAGLE's subgrid physical prescriptions, as these were calibrated at a mass resolution close to that of our simulations. To isolate the impact of AGN feedback, we also perform complementary simulations in which BHs are not seeded, and hence AGN feedback is absent. Detailed descriptions of the EAGLE model (and its recalibration for use at higher resolution) may be found in the EAGLE reference articles listed above, so for brevity we do not include one here. Given that the focus of this work is on the impact of BH growth and AGN feedback, however, we summarise the implementation of these processes in the EAGLE model. BHs are treated as `sink' particles which are seeded on-the-fly at the centres of dark matter haloes when they attain a mass of $10^{10}$ \msol{}$h^{-1}$. They then grow through a combination of BH-BH mergers and Eddington-limited accretion at a rate given by
\begin{equation}
\label{eq:mdotaccr}
    \dot{m}_{\rm accr} = \dot{m}_{\rm Bondi} \times {\rm min}(C_{\rm visc}^{-1}(c_{\rm s}/V_\phi)^3,1),
\end{equation}
where $\dot{m}_{\rm Bondi}$ is the \citet{bondi44} rate,
\begin{equation}
\label{eq:mdotbondi}
    \dot{m}_{\rm Bondi} = \frac{4\pi G^2 M_{\rm BH}^2 \rho}{(c_{\rm s}^2+v^2)^{3/2}},
\end{equation}
Here the quantities $\rho$, $v$, $V_\phi$ and $c_{\rm s}$ describe the properties of the gas within the BH kernel (i.e. within the SPH smoothing length of the BH); $\rho$ is the density, $v$ is the bulk velocity relative to the BH, $V_\phi$ is the rotation velocity around the BH and $c_{\rm s}$ is the sound speed. The factor $C_{\rm visc}^{-1}(c_{\rm s}/V_\phi)^3$ modulates the purely spherically symmetric Bondi rate according to the rotational kinematics of the gas within the BH's SPH kernel, since the accretion of gas with high angular momentum through an accretion disc is posited to proceed more slowly than the Bondi-Hoyle rate \citep[see][]{rosasguevara15}. The value of $C_{\rm visc}$ is set to $2\pi\times 10^3$ in the Recal model employed here\footnote{\citet{rosasguevara15} showed that the results of this prescription are only weakly dependent on the precise value of $C_{\rm visc}$, since the ratio $V_\phi/c_{\rm s}$ is more important.}. Energy from AGN feedback is injected thermally and with a fixed efficiency at a rate $\dot{E}_{\rm AGN}=0.015\dot{m}_{\rm accr}c^2$, by stochastically increasing the temperature of the BH's neighbouring SPH particles by $\Delta T_{\rm AGN}=10^9$ K.


EAGLE's subgrid model includes stochastic elements; the conversion of gas particles to star particles and the coupling of feedback energy to gas require that quasi-random numbers be drawn and compared to probabilities set by the properties of the gas \citep[see][]{schayedv08,dvschaye12}. In a sufficiently large cosmological volume, the emergent properties of the galaxy population are, in a statistical sense, insensitive the the choice of the initial seed for the quasi-random number generator. However, when considering the evolution of individual systems, this choice can cause significant uncertainty \citep[e.g.][]{genel19,keller19,davies21}. To demonstrate that our results are borne out of differences in merger/assembly history and do not simply reflect stochastic noise, we simulate each unique set of initial conditions (both with and without BH seeding and AGN feedback) with nine different seeds each.

\subsection{Producing genetically-modified galaxies}
\label{sec:methods:gm}

We use the genetic modification (GM) technique \citep{roth16,pontzen17} to generate several complementary sets of initial conditions that yield adjusted merger histories for the \organic{} galaxy, whilst preserving the large-scale environment of the system. By applying this technique, the initial conditions generator \textsc{genetIC} \citep{stopyra20} yields a modified field that is as close as possible to the original field and remains consistent with a $\Lambda$CDM cosmology, but which also satisfies constraints that result in a modified merger history.

For our experiment, we focus on a merger that occurs at $t\approx 7$ Gyr ($z\approx 0.74$) for the \organic{} galaxy, with a stellar mass ratio $\mu\equiv M_\star^{\rm infall}/M_\star=0.17$, where $M_\star^{\rm infall}$ and $M_\star$ are the masses of the infalling and primary galaxies respectively. We define `major' mergers to have $\mu>0.25$, and `minor' mergers to have $0.1<\mu<0.25$. This is the galaxy's last significant merger (i.e. $\mu>0.1$), and we selected it for modification because it has the most significant effect on the galaxy's dynamical history; as shown by \citet{davies21}, it removes some rotational support from the galaxy disc, but is not disruptive enough to transform the galaxy into a spheroid by $z=0$. Following the merger, there is a modest increase in the SMBH mass and a decrease in the CGM gas mass, but the central galaxy remains actively star-forming to $z=0$.

We generate a pair of modified initial conditions, designed to adjust the ratio, $\mu$, of this merger, with the aim of enhancing or suppressing its disruptive effects. To achieve this, we identify all particles that are bound to the infalling halo at $z=1.73$, trace them back to their locations in the initial conditions (at $z=99$), and increase or decrease the mean overdensity in the patch of the field containing those particles. To fix the mass of the main galaxy at this time, we apply the simultaneous constraint that the overdensity of the matter that will comprise the main halo at $z=1.73$ remains fixed, and we also require that the overdensity across the Lagrangian region of the $z=0$ halo is kept fixed to preserve the final halo mass. We note that while the galaxy-galaxy merger occurs at $z\approx 0.74$, we adjust the properties of the haloes at $z=1.73$ because this is the latest snapshot output in which the haloes are clearly distinguishable by the SUBFIND algorithm. We henceforth refer to these initial conditions (and the galaxies/haloes that evolve from them) as \textsc{enhanced} and {\sc suppressed}.

For each of our simulations, the state of the particles in the simulation volume is saved in 29 ``snapshot'' outputs between $z=20$ and $z=0$, which match the output times adopted by the main EAGLE simulation suite. To more finely sample the evolution of our candidate galaxy over the course of this merger, we perform a final simulation for each set of initial conditions, outputting 600 snapshots over the time interval $t=1.74-8.35$ Gyr ($z=3.600-0.533$), affording $\approx 10$ Myr time resolution. We hencefore refer to these simulations as ``high-cadence''. These simulations were performed with different seeds to their nine more coarsely-sampled counterparts, such that we did not cherry-pick one particular simulation to examine in detail.

\subsection{Identifying and tracking haloes and galaxies through time}
\label{sec:methods:tangos}

Haloes are tracked on-the-fly in the EAGLE simulations by applying the Friends-of-Friends (FoF) algorithm to the dark matter distribution (with a linking length of 0.2 times the mean interparticle separation), after which gas, star and BH particles are assigned to the FoF group of their nearest dark matter particle. We apply the \textsc{subfind} algorithm \citep{springel01,dolag09} in post-processing in order to identify bound haloes, and use the analysis packages \textsc{pynbody} \citep{pontzen13} and \textsc{tangos} \citep{pontzentremmel18} to track them through time and calculate their properties.

For each simulation, we use \textsc{tangos} to link haloes to their immediate progenitors and descendants based on the number of particles they have in common, yielding a merger tree that allows any halo to be traced back to its progenitors. We identify the main branch of this tree in our fiducial (\organic{}) simulation by computing the sum of the stellar masses along each possible branch, and selecting the branch with the largest value. We then ensure that we reliably identify the \organic{} system's counterpart in our other simulations (evolved from genetically-modified initial conditions and/or with different random number seeds) by finding the halo at $z=5$ that has the greatest number of particles in common with the \organic{} system, and tracing its descendants forwards through time. Since the descendant of a halo is unambiguous, this ensures that we are comparing the same system between different simulations at a given time.

Throughout this work, we calculate the properties of haloes (such as the halo mass, $M_{200}$, and the CGM mass fraction, $f_{\rm CGM}$) within a radius, $r_{200}$, that encloses a mean density equal to 200 times the critical density, centred on the halo centre of mass as determined by \textsc{pynbody} using the shrinking-sphere method \citep{power03}. Galaxy properties, such as the specific star formation rate, are calculated within a 30 pkpc aperture about the centre of mass. We quantify rotational kinematics within spherical apertures using the fraction of kinetic energy that is invested in co-rotational motion,
\begin{equation}
\label{eq:kappa}
    \kappa_{\rm co} = \frac{1}{K}\sum_{i,L_{{\rm z},i}>0}\frac{1}{2}m_i\left(\frac{L_{{\rm z},i}}{m_i R_i}\right)^2,
\end{equation}
where $m_i$ are the masses of particles within the aperture, $L_{{\rm z},i}$ are their angular momenta projected along the rotation axis (the direction of the total angular momentum vector) and $R_i$ are their 2D radial distances from the centre in the plane of the rotation axis. Following \citet{correa17}, the sum is calculated over all particles that are co-rotating (i.e. $L_{{\rm z},i}>0$), and $K$ is the total kinetic energy of all particles in the aperture. We calculate this diagnostic using the publicly available routines of \citet{thob19}.

\section{Results}
\label{sec:results}

We begin our analysis by comparing the evolution of the \organic{} galaxy (and its host halo) to that of its genetically modified counterparts described in Section \ref{sec:methods:gm}, which have been designed to experience either a more (\enhanced{}) or less (\suppressed{}) significant merger at $t\approx 7$ Gyr. Our aim is to determine whether this change to the halo mass accretion history induces strong differences in the growth of the galaxy's SMBH, and leads to a different evolution in the properties of the CGM and galaxy.

\subsection{Mass assembly and merger histories}
\label{sec:results:mass}

\begin{figure}
\includegraphics[width=\columnwidth]{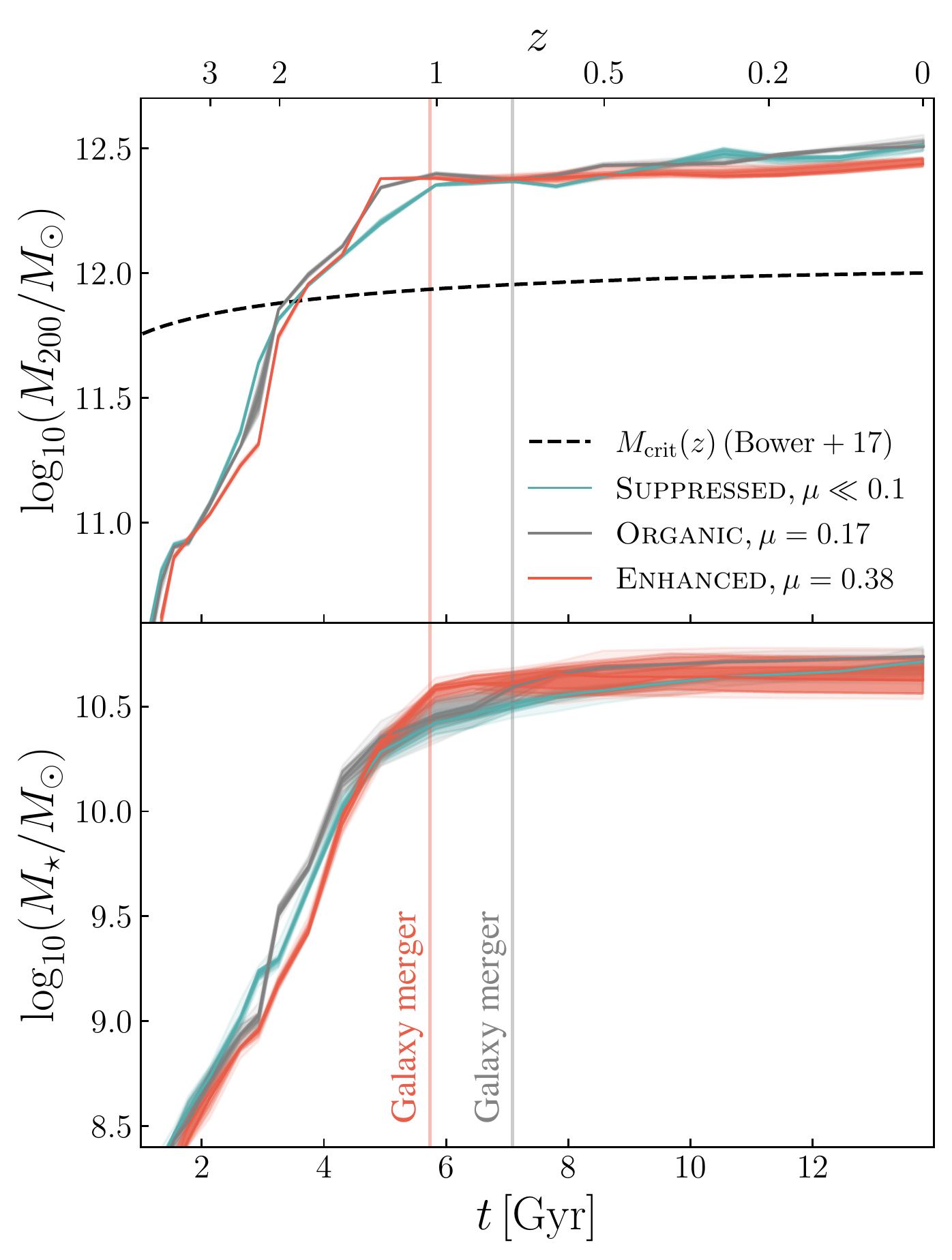}
\caption{The halo mass evolution, $M_{200}(t)$ (upper panel) and stellar mass evolution, $M_\star(t)$ (lower panel) evolution for the \organic{} galaxy and its genetically modified counterparts, in which the target merger has been \enhanced{} or \suppressed{}. Adjusting the merger ratio does not induce strong changes in the halo or stellar mass histories. Shading indicates the distribution of the results over nine simulations, each adopting a different seed for the quasi-random number generator used by EAGLE's stochastic prescriptions for star formation and feedback. Solid lines show the median value at each output time. We mark the times at which the galaxies coalesce in the target merger for the \organic{} and \enhanced{} simulations with vertical lines, and show the redshift-dependent halo mass threshold at which BHs are expected to become able to grow rapidly in EAGLE, $M_{\rm crit}(z)$, with a dashed black line.}
\label{fig:merger:mass}
\end{figure}

We first evaluate the halo and stellar mass accretion histories of the galaxies, to verify that they attain approximately equal present-day masses and to examine the behaviour of the target merger. In Figure \ref{fig:merger:mass} we show, as a function of time, the halo mass ($M_{200}(t)$, upper panel) and stellar mass ($M_\star(t)$, lower panel) of our galaxies. To quantify the uncertainty in these histories that arises from stochasticity in the EAGLE model, we show the results of nine re-simulations of each set of initial conditions, each performed with a different seed for the quasi-random number generator. The median value of the $y$-axis quantity is shown as a function of time with solid lines, and the distribution arising from the nine re-simulations is shown with progressively lighter shading between the pairs of seeds that give increasingly divergent results from the median. We note therefore that any given line does not show the evolution of one particular system, but that the solid lines and shading give the median and distribution of values at a given epoch. Throughout this study, we quantify the scatter at a given time with the interquartile range (IQR); the difference between the third and seventh values in a rank-ordered set of nine values gives a good approximation for this statistic. 

To highlight the times at which our target merger takes place for these simulations, we use the high-cadence simulations to identify when the final coalescence of the galaxies takes place, and highlight this time on Figure \ref{fig:merger:mass} with vertical lines of the appropriate colour for each set of initial conditions. We also use these high-cadence outputs to compute the stellar mass ratio, $\mu$, at the time when the stellar mass of the infalling galaxy is maximal, to account for any stellar mass exchange between the galaxies during close encounters. 

As noted in Section \ref{sec:methods:gm}, the target merger for the \organic{} galaxy has a mass ratio $\mu=0.17$; this merger occurs at $t=7.09$ Gyr. As a result of our genetic modifications, in the \enhanced{} simulations the galaxy experiences a major merger, with a mass ratio $\mu=0.38$. Since the infalling system is more massive than in the \organic{} case, this merger happens at an earlier time, at $t=5.73$ Gyr. In contrast, the infalling galaxy never merges with the main galaxy in the \suppressed{} case, and remains a separate object until $z=0$, at which time $\mu=0.03$. When the host haloes of the progenitor galaxies merge, the infalling galaxy enters the primary halo on a highly tangential orbit, and dynamical friction is insufficient to bring the galaxy to the halo centre. While other very minor mergers occur during the evolution of the \suppressed{} system, the disruptive effect of the target merger has been entirely removed through genetic modification.

The aim of our genetic modifications is to adjust how significantly the galaxy is disrupted by the target merger; this is not solely determined by the stellar mass ratio of the participating systems, but also by the geometry of the merger. \citet{zeng21} showed that star-forming discs in IllustrisTNG can survive even major mergers if the infalling system ``spirals-in" on a near-tangential orbit, while head-on collisions are far more likely to strongly disrupt the disc and yield an elliptical remnant. We find that increasing the mass of the infalling system in the \enhanced{} case causes a more violent, head-on collision; the first interaction of the galaxies has an impact parameter $r_{\rm impact}=0.09r_{200}$, while in the \organic{} case $r_{\rm impact}=0.23r_{200}$. The closest the infalling system ever comes to the main galaxy in the \suppressed{} case is $r_{\rm impact}=0.75r_{200}$. The more disruptive nature of the target merger in the \enhanced{} simulations can therefore be attributed to both an increased stellar mass ratio and a more head-on collision.

By construction, the three sets of initial conditions produce present-day haloes of approximately equal mass; the \organic{}, \enhanced{} and \suppressed{} haloes have masses $M_{200}=10^{12.51}$, $10^{12.44}$ and $10^{12.51}$ M$_\odot$ respectively (IQR$=0.02$, 0.03, 0.01 dex). The final halo masses of the \organic{} and \suppressed{} haloes are consistent, while that of the \enhanced{} halo is slightly lower; we discuss the reason for this in Section \ref{sec:results:bhCGM}. For the majority of the simulation, the halo mass histories track each other closely, but with a notable difference occurring at $t=4.93$ Gyr ($z=1.26$), where the halo masses increase far more rapidly for the \organic{} and \enhanced{} haloes than for the \suppressed{} halo. This is the time at which the haloes of our primary galaxy and the infalling galaxy merge. Our genetic modification experiment is therefore able to control the significance of an individual merger while inducing only very small changes to the overall halo mass accretion history.

In the EAGLE model, SMBHs are expected to grow very little until their host halo reaches a redshift-dependent critical mass, $M_{\rm crit}(z)$. At this mass, the characteristic entropy of the CGM becomes greater than that of outflows driven by stellar feedback, which cease to be buoyant within the halo; this effectively confines gas within the central galaxy and causes the BH to enter a ``rapid growth phase'' \citep{bower17,mcalpine18}. Haloes of the same mass that have different mass assembly histories will cross this threshold at different times, causing large differences in the BH growth history \citep[see][]{davies21}; this is a variation that we must avoid in our experiment, as we would like to isolate the influence of a merger. We show $M_{\rm crit}(z)$ as a dashed black line on the upper panel of Figure \ref{fig:merger:mass}, and confirm that the \organic{} halo and its GM counterparts all cross this threshold at the same time (to an accuracy of $<500$ Myr). 

In the lower panel of Figure \ref{fig:merger:mass} we show the stellar mass evolution of our galaxy and its modified variants. By the present day, the stellar masses of the galaxies are not significantly different, with values of $M_\star=10^{10.74}$, $10^{10.63}$ and $10^{10.71}$ M$_\odot$ (IQR$=0.01$, 0.16, 0.06 dex) for the \organic{}, \enhanced{} and \suppressed{} variants respectively. The stellar mass histories are similar in each case, with significant differences only arising at two points in the system's evolution. The first, at $t\sim 4$ Gyr, occurs because our modifications slightly alter the timing of a major ($\mu \approx 0.5$) merger that occurs early in the system's evolution. The high-cadence simulations reveal that this merger is complete by $t=3.8$ Gyr in the \organic{} case, but later ($t=4.2$ Gyr) in the \suppressed{} case and later still ($t=4.5$ Gyr) in the \enhanced{} case. Small differences such as these are an unavoidable consequence of maintaining consistency with the $\Lambda$CDM correlation function; we will show in later sections that the timing of this merger has a negligible effect on how the SMBH affects the CGM. The second and most notable differences occur at the time of the target merger; the stellar masses of the \enhanced{} and \organic{} galaxies rise rapidly between snapshot outputs as a result of the addition of stars formed ex-situ. The \suppressed{} galaxy does not gain such an ex-situ contribution to its stellar mass since the galaxies never merge; however, this is compensated for by in-situ star formation so that the final stellar masses at $z=0$ are comparable, despite widely differing star formation rates.

\begin{figure}
\includegraphics[width=\columnwidth]{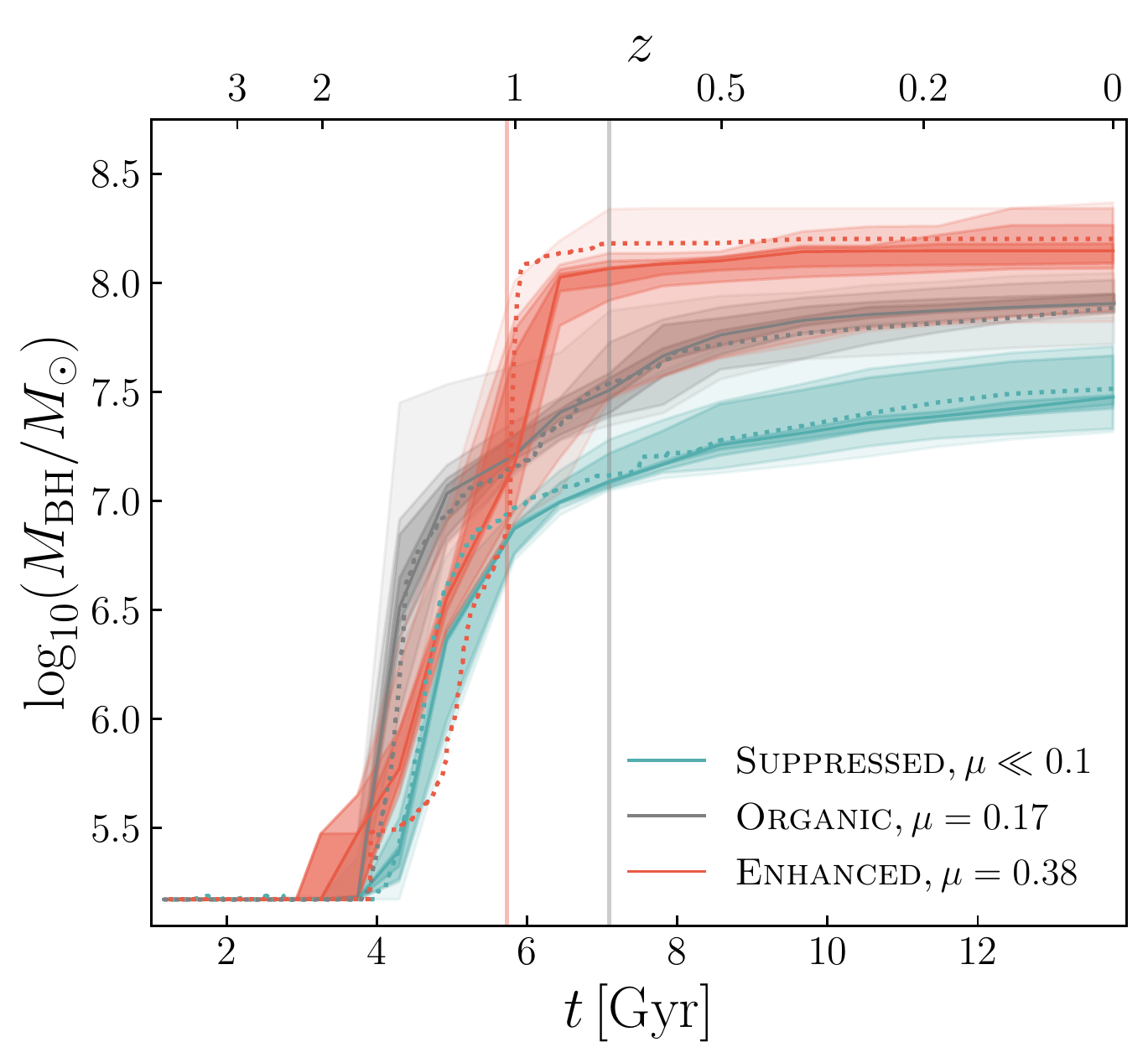}
\caption{Evolution of the central BH mass, $M_{\rm BH}(t)$ for the \organic{} galaxy and its suppressed-merger and enhanced-merger counterparts. We show the evolution in the same fashion as Figure \ref{fig:merger:mass}, starting at $t\approx 4$ Gyr, when the BH begins growing. Dotted lines show the evolution of a `high-cadence' simulation for each set of initial conditions, performed with a unique random number seed and 10 Myr output time resolution. While the three systems exhibit similar star formation histories, adjusting the merger ratio significantly alters the BH growth history; an enhanced merger is followed by rapid BH growth and a more massive BH by $z=0$, while suppressing the merger yields less integrated growth.}
\label{fig:merger:bh}
\end{figure}

\subsection{BH growth and evolution of the CGM mass fraction}
\label{sec:results:bhCGM}

We now examine how adjusting the merger ratio affects the evolution of the SMBH and its effect on the CGM. Figure \ref{fig:merger:bh} shows, in the same fashion as Figure \ref{fig:merger:mass}, the evolution of the central SMBH mass, $M_{\rm BH}$, for the \organic{}, \enhanced{} and \suppressed{} galaxies. The evolution of $M_{\rm BH}$ is also overlaid for the three high-cadence simulations with dotted lines. 

While the halo and stellar mass histories of our galaxy are only mildly affected by our modifications, enhancing or suppressing the target merger induces dramatic differences in the growth of the central SMBH. The \organic{} BH grows steadily throughout the target merger event, and attains a median final mass $M_{\rm BH}=10^{7.9}$ \msol{} (IQR$=0.1$ dex). The \enhanced{} BH initially has a lower mass than its \organic{} counterpart, but grows very rapidly following the merger and attains a greater median final mass $M_{\rm BH}=10^{8.2}$ \msol{} (IQR$=0.2$ dex). The BH in the \suppressed{} galaxy, for which the target merger does not occur, grows at a slower rate than its \organic{} counterpart from $t\approx 6$ Gyr onwards, and attains a lower median final mass of $10^{7.5}$ \msol{} (IQR$=0.2$ dex). We note that differences in the timing of the earlier merger at $t\approx 4$ Gyr causes minor variations in the subsequent evolution of $M_{\rm BH}$, as was the case for the stellar mass (Section \ref{sec:results:mass}). However, these differences are very small in comparison to those induced by the target merger.

The BH growth histories exhibit significantly more stochastic scatter than the halo and stellar mass histories. The growth rate of the BH depends on the availability of gas, which in part depends on both the conversion rate of gas into stars and on the injection of energy by stellar and AGN feedback, processes that are implemented stochastically in the EAGLE model. The deterministic changes to $M_{\rm BH}$ in response to our adjustment of the target merger are, however, significantly larger than this scatter. We have therefore demonstrated that merger events can modulate the total energy injected through AGN feedback (which is proportional to the total mass accreted onto the BH), and change how rapidly large fractions of that energy are injected into the system.

\begin{figure}
\includegraphics[width=\columnwidth]{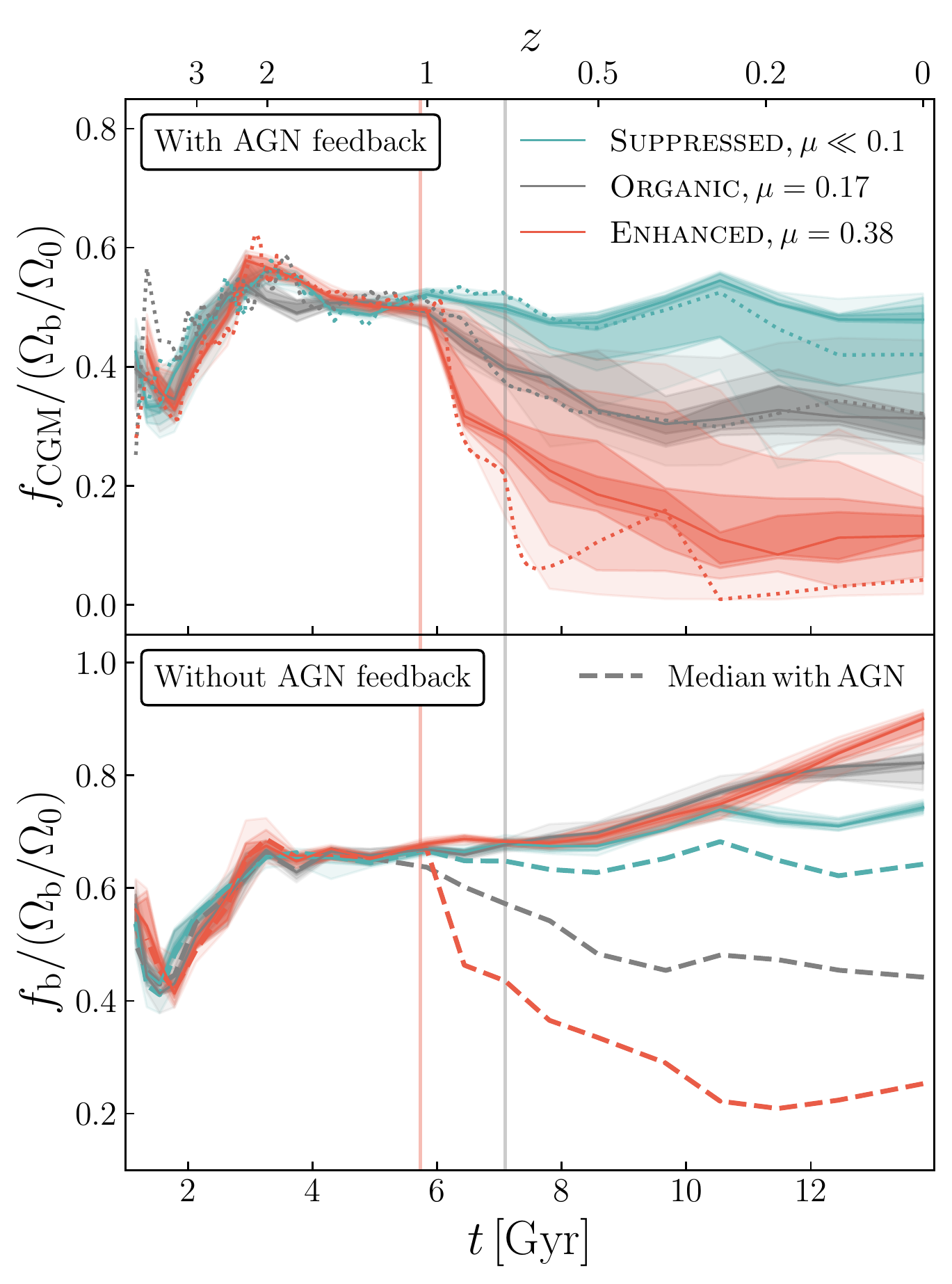}
\caption{Upper panel: evolution of the CGM mass fraction, $f_{\rm CGM}(t)$, for the \organic{}, \enhanced{} and \suppressed{} simulations, shown in the same fashion as Figure \ref{fig:merger:bh}. Lower panel: solid lines and shading indicate the evolution of the halo baryon fraction, $f_{\rm b}(t)$ in complementary simulations without AGN feedback, and dashed lines indicate the median $f_{\rm b}(t)$ in simulations with AGN feedback. Both $f_{\rm CGM}(t)$ and $f_{\rm b}(t)$ are normalised by the cosmic average baryon fraction, $\Omega_{\rm b}/\Omega_0$. Relative to the \organic{} system, the rapid growth of a more massive BH in the \enhanced{} system depletes the CGM of a greater fraction of its baryons, while the steadier growth of a lower-mass BH in the \suppressed{} system leads to the expulsion of fewer baryons. In the absence of AGN feedback, the halo is not depleted of baryons in any simulation.} 
\label{fig:merger:cgm}
\end{figure}

The energy injected by AGN feedback can drive strong changes in the gas content of the CGM, which we quantify with the CGM mass fraction, $f_{\rm CGM}$. In the upper panel of Figure \ref{fig:merger:cgm} we show the evolution of $f_{\rm CGM}$, normalised to the cosmic baryon fraction, $\Omega_{\rm b}/\Omega_0$, for our three families of simulations in the same fashion as Figure \ref{fig:merger:bh}. Prior to the enhanced target merger, the CGM mass fractions of all three systems are approximately equal at $f_{\rm CGM}\approx 0.50$, but then diverge from this value. Since we have confirmed that the three systems form approximately equal stellar masses, these differences in the CGM content must be due to the expulsion of baryons from the halo (and, additionally, the prevention of baryons from entering the halo) and are not the result of differences in how much of the CGM cools and forms stars.

The differences in the evolution of \gasfrac{} reflect the differences in the BH growth history for our three systems; the CGM in the \enhanced{} system is depleted of a larger fraction of its baryons than the CGM in the \organic{} system, while the \suppressed{} system retains a greater fraction of its CGM mass. The $z=0$ median CGM mass fractions are significantly different for the three systems; they are $f_{\rm CGM}=0.48$ (\suppressed{}, IQR$=0.11$), 0.31 (\organic{}, IQR$=0.05$) and 0.12 (\enhanced{}, IQR$=0.07$). The distributions of $f_{\rm CGM}(t)$ show that the stochastic variations in $M_{\rm BH}(t)$ correspond to large uncertainty in the CGM mass fraction; however the distributions are distinctly different after the median values diverge, with some overlap between only the most extreme values for each system. The reduction of $f_{\rm CGM}$ in the \enhanced{} system is the reason why its halo exhibits a slightly lower final halo mass (Figure \ref{fig:merger:mass}); gas expulsion at high redshift both reduces the halo gas mass and suppresses the long-term accretion of baryons onto the halo \citep[e.g.][]{wright20}. 

In the \enhanced{} system, the rapid BH growth induced by the target (major) merger triggers the expulsion of gas from the CGM, since $f_{\rm CGM}$ is approximately constant prior to the merger, but declines sharply immediately after the galaxies coalesce. In comparison, the evolution of $f_{\rm CGM}(t)$ and $M_{\rm BH}(t)$ for the \organic{} system suggest that a less significant target merger has little effect; when it occurs $f_{\rm CGM}$ is already declining steadily, and this decline does not accelerate following the merger. However, prior to the target merger its BH is significantly more massive than its counterpart in the \suppressed{} system, for which the merger is absent, indicating that a minor merger can drive slow enhancement of SMBH growth long before the merger is completed. As a result of this difference, \gasfrac{} declines gradually for the \organic{} system, but remains constant in the absence of a merger in the \suppressed{} system.

To confirm that adjusting the merger ratio changes the evolution of $f_{\rm CGM}$ because of differences in the expulsion of gas by AGN feedback, we examine our simulations in which BHs are not seeded and hence AGN feedback is absent. To isolate the expulsive effect of AGN feedback, we examine the evolution of the halo {\it baryon} fraction ($f_{\rm b} \equiv M_{\rm b}/M_{200}$, where $M_{\rm b}$ is the total mass of all baryons within $r_{200}$) rather than the CGM mass fraction, since disabling AGN feedback can allow a larger fraction of the CGM to form stars. The evolution of $f_{\rm b}(t)$ (normalised to the cosmic baryon fraction) in the absence of AGN feedback is shown for our three systems in the lower panel of Figure \ref{fig:merger:cgm}, and the median $f_{\rm b}$ for simulations including AGN feedback are shown with dashed lines for comparison. Regardless of merger ratio, the halo is not depleted of baryons following the target merger, and $f_{\rm b}$ increases as the halo potential deepens. The inclusion of AGN feedback expels a greater fraction of the halo baryons as the merger ratio increases; with AGN feedback the \suppressed{} halo retains 86\% of the baryons that it contains without AGN, while the \enhanced{} halo retains only 28\%.

A Milky Way-like galaxy may experience several major or minor mergers over its lifetime, though the majority of these will occur at early times ($z\gtrsim 2$, see e.g. \citealt{rodriguezgomez15}). Mergers at such early times will likely not induce AGN-driven outflows in EAGLE galaxies similar to our \organic{} galaxy, since $M_{200}<M_{\rm crit}$ at these times (see Section \ref{sec:results:mass}) for typical halo assembly histories. All three systems experience a major merger at $t=3.8-4.2$ Gyr, shortly after $M_{\rm crit}$ is exceeded, and the difference in the timing of this merger (see Section \ref{sec:results:mass}) may explain why the BHs begin growing at slightly different times (Figure \ref{fig:merger:bh}). However, the CGM appears insensitive to these early differences in the BH's growth (Figure \ref{fig:merger:cgm}) because the BH mass is low, and therefore so is the AGN energy injection rate ($\dot{E}_{\rm AGN} \propto \dot{m}_{\rm accr} \propto M_{\rm BH}^2$, see Section \ref{sec:methods:ics}). The target merger is the only significant merger (i.e. $\mu>0.1$) that occurs when sufficiently high energy injection rates can be reached to expel gas from the CGM. By systematically adjusting the stellar mass ratio of this merger we can show clearly that a disruptive merger can lead to strong CGM expulsion, while preventing a disruptive event from occurring can inhibit the growth of the BH.

\subsection{Effect of mergers on the black hole accretion history}
\label{sec:results:vicinity}

We next explore why adjusting the mass ratio of a merger impacts the growth of the central BH by studying the evolution of the gas in the its vicinity using the high-cadence simulations. The dotted lines on Figures \ref{fig:merger:bh} and \ref{fig:merger:cgm} show that the evolution of $M_{\rm BH}$ and $f_{\rm CGM}$ for the galaxies in these simulations falls within the scatter induced by stochasticity, and that these simulations are consistent with those adpoting our fiducial output frequency.

\begin{figure}
\includegraphics[width=\columnwidth]{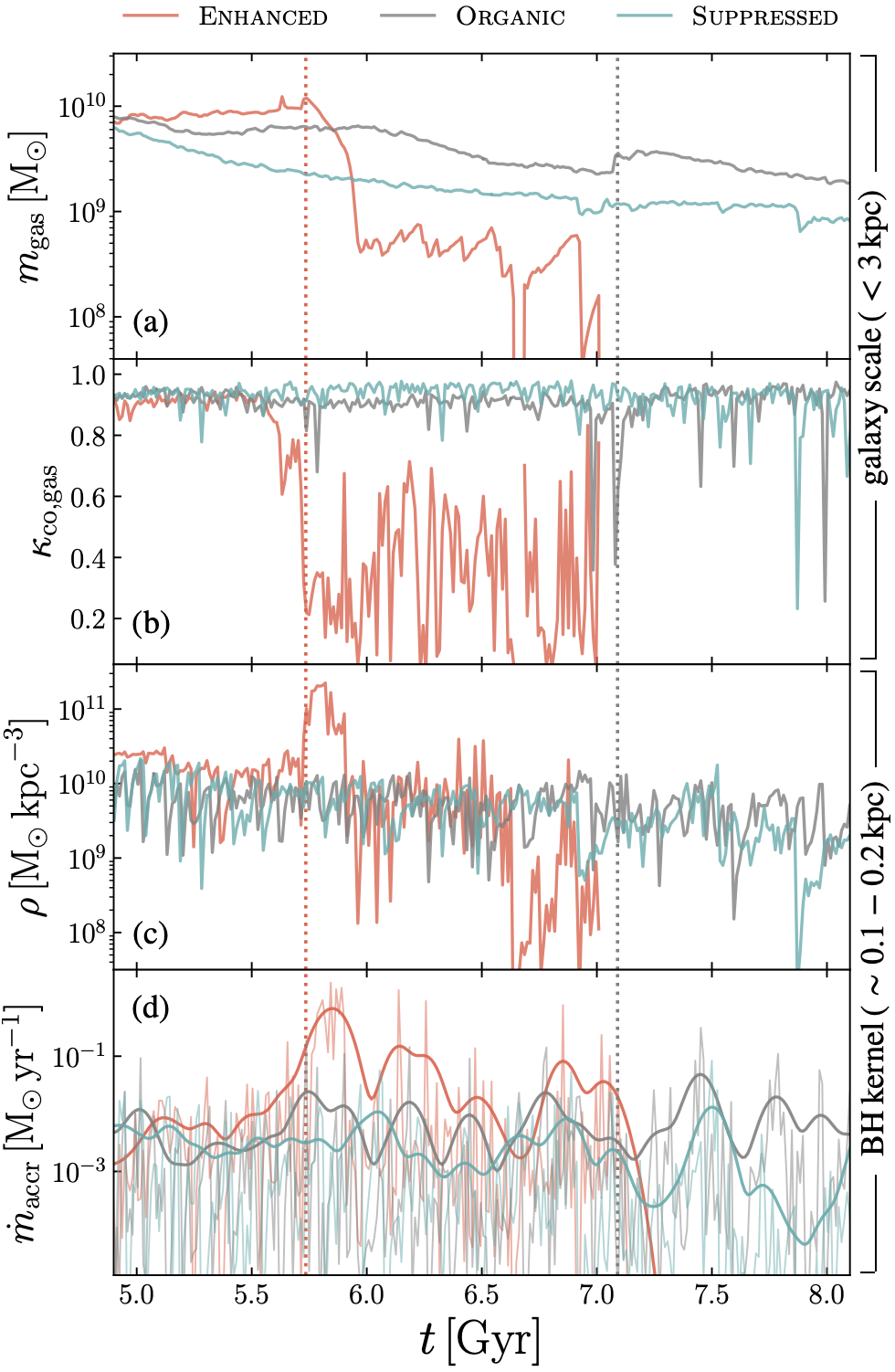}  
\caption{Evolution of (a) the mass, $m_{\rm gas}$, and (b) co-rotational kinetic energy fraction, $\kappa_{\rm co,gas}$, of the gas within 3 pkpc of the BHs in our high-cadence simulations. Panels (c) and (d) show the evolution of the SPH kernel densities, $\rho$, and accretion rates, $\dot{m}_{\rm accr}$, of the BHs respectively. We focus on the time period over which the target mergers occur, which are highlighted with vertical dotted lines. The enhanced merger is followed by a period of very rapid BH accretion, because the gas loses rotational support and collapses to the galaxy centre, greatly increasing the density around the BH. The less significant merger experienced by the \organic{} galaxy does not remove this rotational support, and no such rapid BH accretion occurs.}
\label{fig:vicinity}
\end{figure}

In Figure \ref{fig:vicinity} we show the evolution of several properties of the gas in the vicinity of the BH for the \organic{}, \enhanced{} and \suppressed{} galaxies over the interval $t=5-8$ Gyr, during which the target mergers occur, as indicated by vertical dotted lines. In panels (a) and (b) we show properties measured within a spherical aperture of radius 3 pkpc about the position of the BH, which we consider to be representative of the galaxy scale gas distribution; the total gas mass, $m_{\rm gas}$ and the fraction of the kinetic energy of the gas that is invested in co-rotational motion, $\kappa_{\rm co,gas}$, calculated using Equation \ref{eq:kappa}. We do not show results when there are fewer than 10 particles within this aperture. In panel (c) we show the density $\rho$ within the BH kernel, which is used in the calculation of the accretion rate of the BH, $\dot{m}_{\rm accr}$ (see Section \ref{sec:methods:ics}). The evolution of $\dot{m}_{\rm accr}$ is shown in panel (d); since it varies significantly over short timescales, for clarity we smooth it with a Gaussian kernel of standard deviation 40 Myr and show the result with solid lines, whilst showing the instantaneous accretion rate with thinner, fainter lines. Through examination of these quantities, we can investigate how changes to the properties of the larger-scale gas distribution affect the conditions within the BH kernel and modulate the BH accretion rate.

We focus first on the \enhanced{} galaxy, in which the BH grows rapidly following the merger. Panel (a) shows that $m_{\rm gas}$ does not significantly increase at the time of the merger, indicating that rapid BH growth does not occur because a large supply of gas is introduced to the central region of the galaxy. However, the kinematics of the gas already present within 3 pkpc of the BH are transformed by the gravitational effects of the merger, as shown by the sharp decline in the $\kappa_{\rm co,gas}$ diagnostic in panel (b). Prior to the merger (at $t=5.5$ Gyr), the gas forms a near-perfectly co-rotating disc, with $\approx 90\%$ of its kinetic energy invested in co-rotation, but this fraction drops to $\approx 20$\% after the merger takes place. Now lacking rotational support, the gas on ``galaxy scales'' flows to the centre of the galaxy and towards the smoothing kernel of the BH, which has a radius $h_{\rm BH}\approx 0.1-0.2$ pkpc over the course of the merger. As seen in Panel (c), the smoothed gas density within the kernel increases sharply as this occurs (by a factor $\approx 15$) and remains elevated until $t=5.9$ Gyr. This increase causes a commensurate increase in the BH accretion rate, as shown in panel (d), initiating the expulsion of the CGM by AGN feedback.

\begin{figure*}
\includegraphics[width=\textwidth]{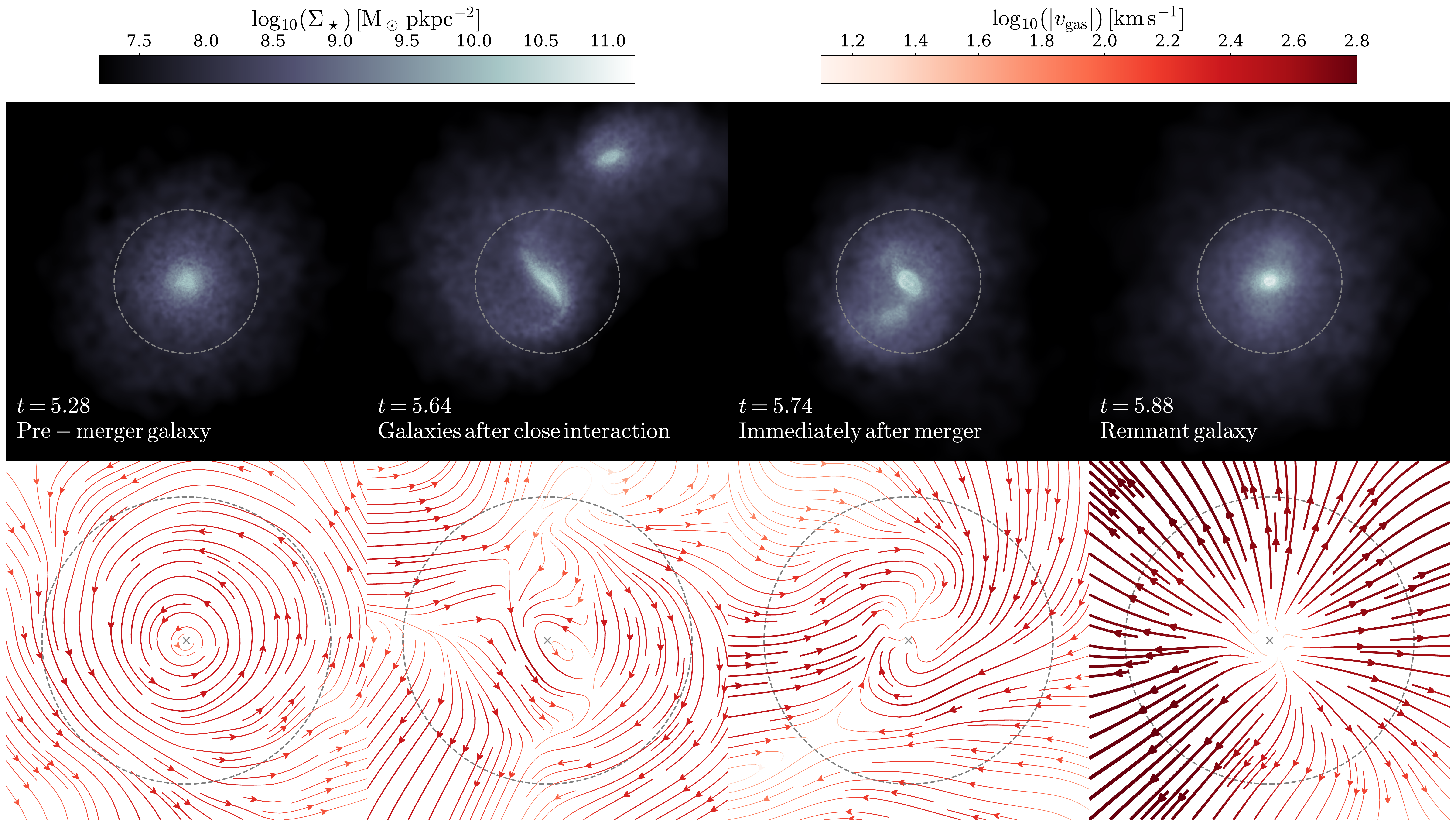}
\caption{Illustration of how ordered co-rotational gas motion around the central SMBH of the \enhanced{} galaxy is disrupted over the course of the target merger. Images in the upper row show the stellar surface density, $\Sigma_\star$, of the \enhanced{} galaxy in a 15 pkpc wide frame. Streamlines in the lower row show the mass-weighted mean gas velocity in a 1 pkpc deep slice, zoomed in by a factor of 2 relative to the upper row. Streamline colours and widths indicate the magnitude, $|v_{\rm gas}|$, of the mass-weighted mean gas velocity. All panels are centred on the location of the SMBH (indicated with a grey cross in the lower row), and are aligned consistently such that the galaxy disc in the first image (i.e. at $t=5.28$ Gyr) is viewed face-on. The spherical aperture of radius 3 pkpc used in Figure \ref{fig:vicinity} is shown with a dashed line. The co-rotational motion of gas in the galaxy disc prior to the merger (first column) is disrupted as the merging galaxies interact (second column), yielding rapid inflow towards the SMBH when the merger is complete (third column) and driving AGN feedback (fourth column).}
\label{fig:streams}
\end{figure*}

Following the merger, the majority of the gas mass in the vicinity of the BH is either accreted or ejected through AGN feedback, and rapid accretion onto the BH ends when $\rho$ sharply declines to below its pre-merger value. After this time, the \kappacogas{} diagnostic indicates a turbulent spheroid of gas with little rotational support. Several very brief periods of high $\dot{m}_{\rm accr}$ occur while the galaxy is in this state, until a final AGN feedback event at $t\sim 7$ Gyr leaves the galaxy very gas-poor.

In the EAGLE model, the accretion rate is also modulated by the gas vorticity within the BH kernel, $V_\phi/c_{\rm s}$ (see Section \ref{sec:methods:ics}). However, we find that unlike the rotational motion of gas on larger scales, the vorticity within the BH kernel is not significantly affected by the merger. The BH accretion rate is therefore responding to a change in $\rho$ within the kernel and not to a change in the gas vorticity. This result is important for the generalisation of our findings, as it demonstrates that they do not originate from the specifics of EAGLE's BH growth prescription. Instead, the rapid accretion onto the BH occurs because of the disruption of gas on the scale of the resolved galaxy disc, driving high densities at the galactic centre. This process is likely common to cosmological simulations performed with other models.

We illustrate this sequence of events in Figure \ref{fig:streams}, in which we show maps of the stellar surface density ($\Sigma_\star$, upper row) and gas velocity (lower row) in the \enhanced{} galaxy at four output times over the course of the merger. Streamlines in the lower row show the mass-weighted mean gas velocity in a 1 pkpc deep slice, and their colours and widths indicate the magnitude of this velocity, $|v_{\rm gas}|$. We show stellar surface density maps with a 15 pkpc field of view, and zoom in by a factor of 2 for the gas velocity maps. All panels are centred on the location of the SMBH (indicated with a grey cross in the lower row), and are aligned consistently such that the galaxy disc in the first image (i.e. at $t=5.28$ Gyr) is viewed face-on. We indicate a spherical aperture of radius 3 pkpc about the SMBH with a dashed line. 

Prior to the merger (first column, at $t=5.28$ Gyr) the galaxy is a star-forming disc in which gas is co-rotating, with little to no inflow towards the BH. The galaxy then experiences a close interaction with the infalling system (with $r_{\rm impact}=2.7$ kpc); the second column shows the pair of galaxies 10 Myr after this interaction, revealing that both the stellar disc of the main galaxy and the motion of gas within it are strongly disrupted. The galaxies merge 100 Myr later; the third column shows the system immediately after this happens. The gas within 3 pkpc of the BH is now predominantly falling inwards at high velocity, fuelling BH growth and giving rise to AGN feedback, which drives outflows from the remnant galaxy as seen in the fourth column. We show the magnitude of the gas velocities for illustrative purposes, and refer the reader to \citet{mitchell20} for a detailed study of feedback-driven outflows in the EAGLE model.

We now turn our attention to the \organic{} galaxy, represented by grey lines on Figure \ref{fig:vicinity}. The minor merger it experiences (at the time indicated by the vertical grey dotted line) does bring new gas within 3 pkpc of the BH, and $m_{\rm gas}$ increases by a factor 1.7 between $t=7.00-7.19$ Gyr. Critically, however, the co-rotational motion of the gas in the \organic{} galaxy is not disrupted by this more gentle merger. Within the 3 pkpc aperture, $\kappa_{\rm co,gas}$ is characteristic of near-perfect co-rotational motion prior to the merger, briefly dips as the galaxies coalesce, then quickly returns to its pre-merger value. There is therefore no sudden inflow of gas towards the BH, and no significant increase in $\rho$ or the accretion rate. The BH in the \organic{} galaxy grows steadily over the course of the merger, and as shown by Figure \ref{fig:merger:cgm}, the AGN feedback associated with this growth is less efficient at removing baryons from the CGM than the intense, brief episodes of feedback experienced by the \enhanced{} galaxy.

The target merger in the \organic{} system does appear to cause subtle, long-term effects that are revealed through comparison with the \suppressed{} system, for which the target merger never occurs. Both systems host co-rotating discs of gas over the time period shown in Figure \ref{fig:vicinity}, with similar $\kappa_{\rm co,gas}$, however there is more gas present within the inner 3 pkpc of the \organic{} system. The tidal forces exerted on the main galaxy by the merging system as it falls to the centre of the halo may cause gas to migrate to the centre of the galaxy, yielding a higher $m_{\rm gas}$ in the inner 3 pkpc than is found in the absence of any merging activity. Note that this is a far more gradual and subtle effect than the sudden collapse of gas towards the BH seen in the \enhanced{} galaxy. This larger reservoir of gas allows the \organic{} BH to accrete more gas (and hence, inject more feedback energy) over the lifetime of the system. It therefore forms a more massive BH and ejects more gas from the CGM (Figures \ref{fig:merger:bh} and \ref{fig:merger:cgm}), than its counterpart in the \suppressed{} galaxy. 

It appears that mergers can unlock the AGN-driven expulsion of the CGM. The small increase in $m_{\rm gas}$ following the enhanced merger indicates that the gas required to drive a strong outflow was already present in the vicinity of the BH, but could not be accreted due to its high angular momentum. The disruptive effects of a merger can bring this gas close to the BH, and this can happen very rapidly following a disruptive merger, or more gradually as a result of tidal interactions. We note that as shown in Figure \ref{fig:merger:cgm}, $f_{\rm CGM}$ continues to fall many Gyr after the merger unlocks the AGN in the \enhanced{} galaxy, and $M_{\rm BH}$ continues to slowly rise. This likely occurs because the turbulent remnant galaxy is able to rapidly remove angular momentum from any gas reaching the system, which will then accrete onto the BH rather than forming stars \citep[see also][]{pontzen17}, driving intermittent AGN feedback.

\subsection{Influence of merger-induced AGN feedback on CGM cooling}
\label{sec:results:preventative}

\begin{figure}
\includegraphics[width=\columnwidth]{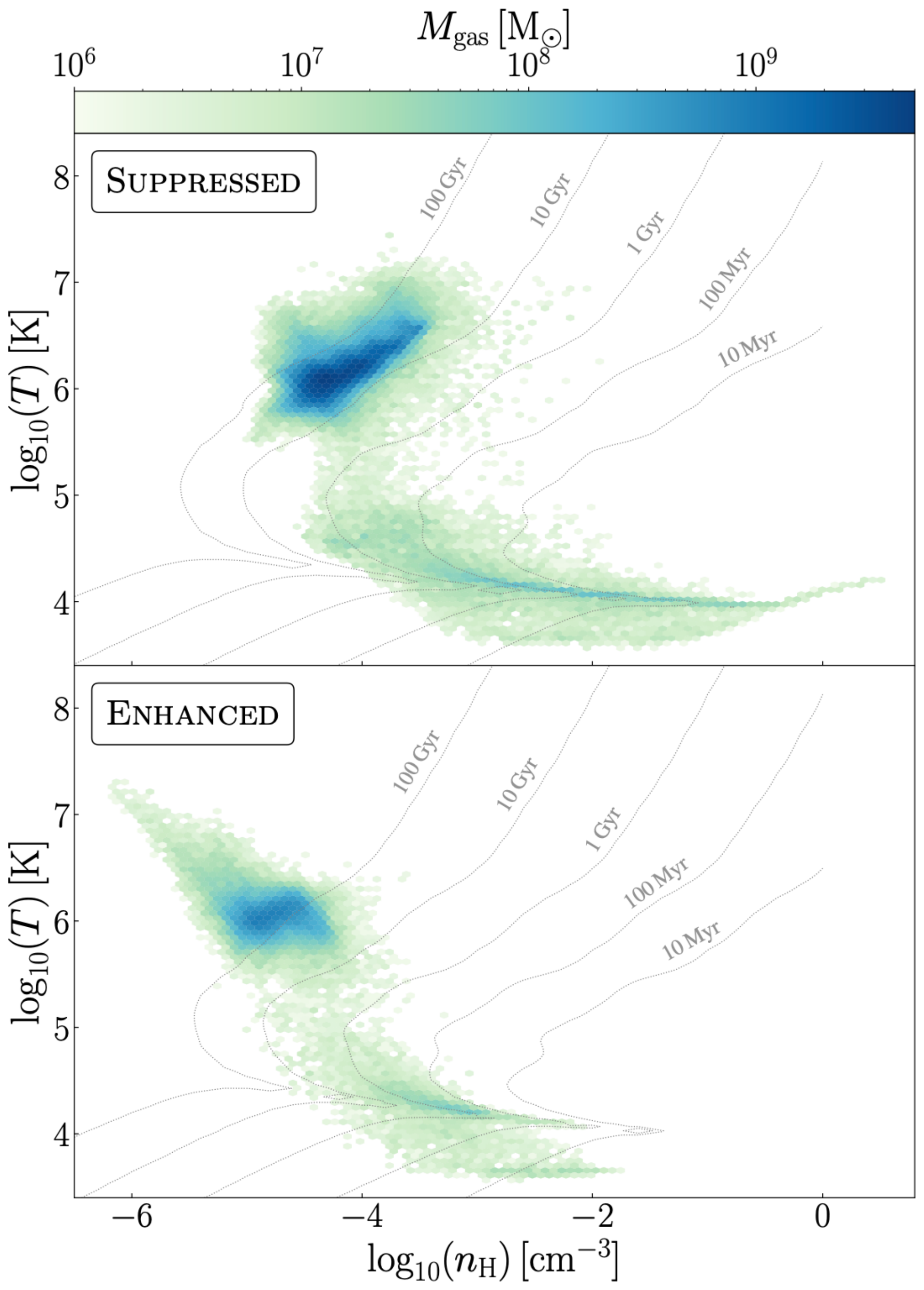}  
\caption{Present-day temperature-density histograms for the gas in the CGM of our \suppressed{} (upper) and \enhanced{} (lower) systems. Distributions are shown for the system with the median final $f_{\rm CGM}$ within the seed-to-seed scatter. Isocontours of constant cooling time are overlaid, using the mean metallicity and He/H ratio of the halo in each simulation. The expulsion of a large fraction of the CGM in the \enhanced{} system shifts the gas remaining in the halo to lower densities and slightly higher temperatures, elevating its cooling time.}
\label{fig:phase}
\end{figure}

Having established that the AGN feedback triggered by a disruptive merger is capable of ejecting a significant fraction of the CGM from the halo, we now examine how the physical state of the remaining CGM is modified by this process. In Figure \ref{fig:phase} we show mass-weighted distributions of gas in the $z=0$ CGM of the \suppressed{} and \enhanced{} galaxies (upper and lower panels respectively) in temperature-density space. Specifically, we show distributions of the temperature, $T$, versus the number density of hydrogen atoms, $n_{\rm H}$, for all non-star forming gas within the virial radius, $r_{200}$. Note that the distributions are not normalised to the total mass of the CGM; differences in the CGM mass (and hence $f_{\rm CGM}$, since the haloes have similar $M_{200}$) are reflected in the distribution amplitudes. In each case, we show distributions for the simulation that has the median final $f_{\rm CGM}$. Isocontours of constant radiative cooling (or, at low $n_{\rm H}$ and $T$, heating) time, obtained from the CLOUDY lookup tables employed by the EAGLE model, are overlaid. For simplicity, we assume solar abundance ratios for the metallic elements but rescale the cooling function to the mean metallicity for each distribution following \citet[][their Equation 5]{wiersma09}.

The impact that the AGN feedback induced by a merger can have on the physical state of the CGM is evident from the distributions shown in Figure \ref{fig:phase}. The \enhanced{} CGM has been significantly depleted of gas, and the majority of the mass that remains has been shifted to both lower densities and slightly higher temperatures. As a result, the cooling time of the gas has been extended, as can be seen with reference to the isocontours of constant cooling time. In the \suppressed{} system, $6.5\times 10^{10}$ \msol{} of gas can radiatively cool within a Hubble time, $t_{\rm H}$, while only $5.7\times 10^{9}$ \msol{} of gas in the \enhanced{} system can do so. In our simulations in which AGN feedback is absent, the temperature-density distribution of the gas is insensitive to the target merger, and resembles that of the \suppressed{} system in Figure \ref{fig:phase}.

In order to continue forming stars, central star-forming galaxies must replenish their interstellar medium with gas from the CGM, and this process will be inhibited in systems such as our \enhanced{} galaxy where the CGM cooling time is elevated through expulsion. This is a form of preventative feedback, and has been identified in earlier statistical studies of simulated galaxy populations as a route to the quenching of star formation in central galaxies \citep[e.g.][]{davies20,zinger20}. The characteristic radiative cooling time of the CGM likely increases primarily due to the reduction in the density of the CGM (since the radiative cooling time $t_{\rm cool}\propto 1/n_{\rm H}^2$) with the increase in temperature being a secondary effect (as $t_{\rm cool}\propto 1/T$). The shift in the density seen in Figure \ref{fig:phase} is substantially greater than the shift in temperature, in agreement with the findings of \citet{davies21}, where we showed that in the EAGLE model, AGN feedback does not strongly alter the CGM temperature profile, but reconfigures the CGM at a lower density without affecting the shape of the density profile. Here we have shown that this can be achieved by increasing the significance of an individual merger.

\subsection{How merger-driven CGM transformation affects the galaxy}
\label{sec:results:galaxy}

We conclude our study by examining how the transformation of the baryon cycle following a disruptive merger affects the star formation history (SFH) and kinematics of the central galaxy. We quantify the SFH by calculating the evolution of the specific star formation rate, sSFR$=\dot{M}_\star/M_\star$, integrated over the preceding 300 Myr at each output time\footnote{The results are not strongly sensitive to this choice, and we have verified that choosing an integration time of 100 Myr or 1 Gyr yields very similar results.}. We quantify the kinematics of the system with the co-rotational kinetic energy fraction, \kappacostar{}, of the stellar component of the galaxy. We show the time evolution of these quantities in Figures \ref{fig:merger:ssfr} and \ref{fig:merger:kappa} in the same fashion as Figure \ref{fig:merger:cgm}, imposing a minimum sSFR of $10^{-13}$ yr$^{-1}$ for clarity. 

In Figure \ref{fig:merger:ssfr} we also show the location of the green valley as a function of time for galaxies of a comparable stellar mass in the largest EAGLE simulation volume \citep[Ref-L100N1504,][]{schaye15,crain15}. To obtain this at each output time, we use the publicly-available EAGLE galaxy catalogues \citep{mcalpine16} to find the sSFR of all galaxies in a 0.2 dex-wide window about the current stellar mass of the \organic{} galaxy. Following \citet{wright19}, we then take the locus of the star-forming main sequence to be the mean sSFR of these galaxies, subject to a threshold $\log_{10}({\rm sSFR/yr^{-1}})>-11+0.5z$, and define the green valley to lie within 5\% and 50\% of this value.

\subsubsection{Star formation activity}

\begin{figure}
\includegraphics[width=\columnwidth]{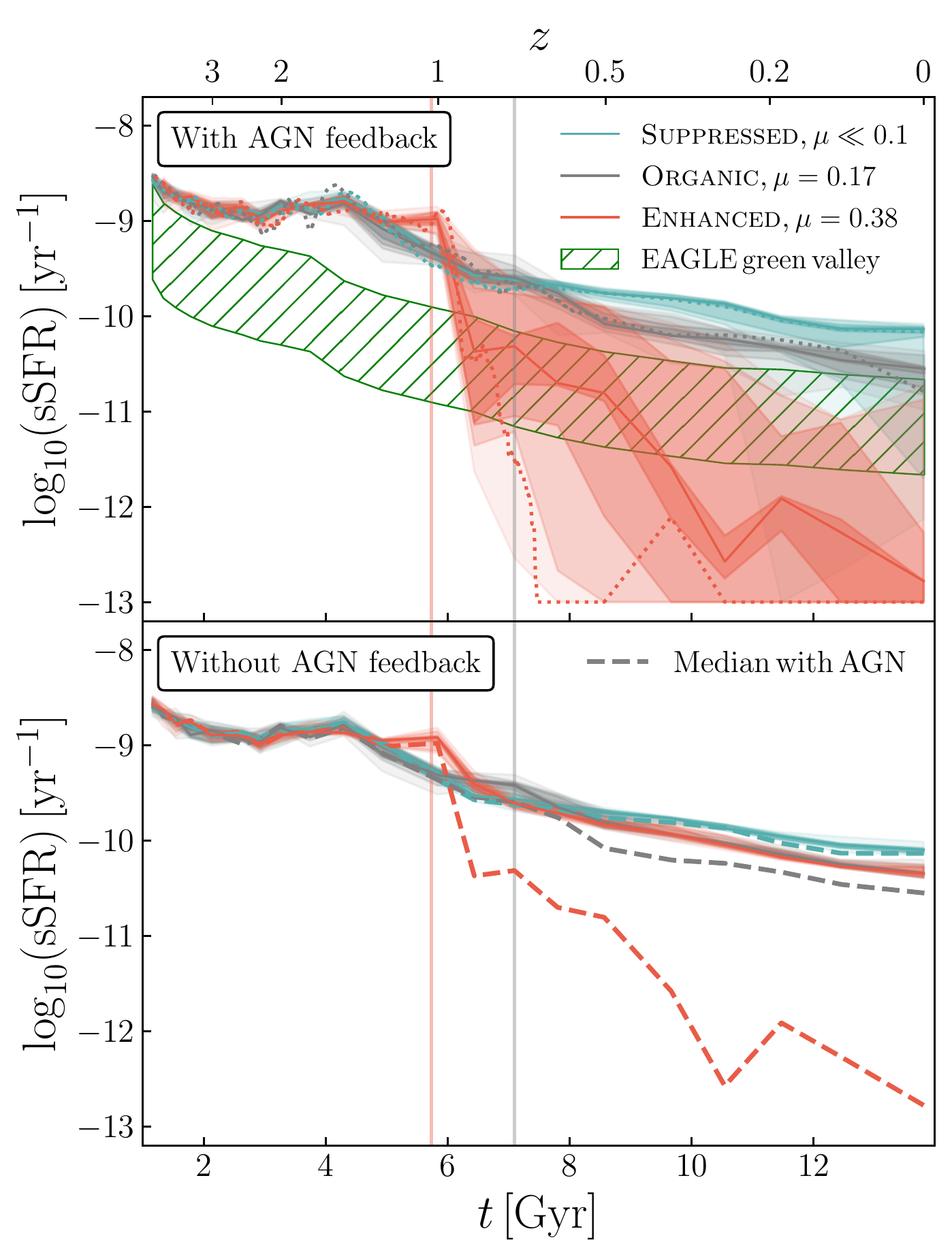}
\caption{Evolution of the specific star formation rate (sSFR) of the \organic{} galaxy and its suppressed-merger and enhanced-merger counterparts, shown in the same fashion as Figure \ref{fig:merger:cgm}. Enhancing the merger, which causes a more significant elevation of the CGM cooling time by AGN feedback, causes the galaxy to quench, whereas it remains actively star-forming at the present day in the \organic{} and \suppressed{} simulations (upper panel). In the absence of AGN feedback, all galaxies remain star-forming (lower panel).}
\label{fig:merger:ssfr}
\end{figure}

The upper panel of Figure \ref{fig:merger:ssfr} shows that adjustment of the target merger ratio leads to significant changes in the SFH of the galaxy. The SFHs of the \enhanced{} and \organic{} galaxies diverge from that of the \suppressed{} galaxy when their target mergers occur. The \suppressed{} galaxy remains on the star-forming main sequence until $z=0$, with a final median sSFR of $10^{-10.14}$ yr$^{-1}$ (IQR=$0.10$). Following the merger, the median sSFR of the \organic{} galaxy falls $\sim 0.3$ dex below that of the \suppressed{} galaxy, and its final median sSFR of $=10^{-10.55}$ yr$^{-1}$ (IQR=$0.22$) is indicative of a galaxy moving into the green valley. The more disruptive merger experienced by the \enhanced{} galaxy has a more dramatic impact on its SFH; the median sSFR is briefly enhanced relative to the other systems immediately after the merger, likely due to a short-lived increase in the central gas density, but then falls rapidly into the green valley as the interstellar medium is ejected by AGN feedback. The median sSFR remains in the green valley for $\approx 3$ Gyr before falling below it, indicating that the galaxy is quenched until $z=0$. The SFHs in resimulations of these galaxies without AGN feedback, shown in the lower panel of Figure \ref{fig:merger:ssfr}, demonstrate that AGN feedback is responsible for these differences. When no BHs are seeded in the galaxies, the enhanced merger does not lead to quenching, and the SFHs and final sSFRs of all three galaxies remain similar.

While the quenching of the \enhanced{} galaxy is undoubtedly caused by the merger that it experiences, it is important to note that in the short term, the AGN feedback induced by the merger is typically only sufficient to drive the galaxy into the green valley, and not to quench it entirely. The galaxy continues forming stars at a slower rate with its remaining interstellar gas, though the sSFR continues to decline because the baryon cycle has been fundamentally altered by AGN feedback as discussed in the previous section. The `refuelling' of the galaxy is inhibited because very little gas remains in the CGM that can cool efficiently. Once the remaining interstellar gas is consumed by star formation and/or BH growth, or ejected by feedback, the galaxy leaves the green valley, quenches, and remains quiescent over long timescales. The time a post-merger galaxy takes to pass through the green valley in this way is likely sensitive to many factors, though the evolution of the \enhanced{} system here indicates that quenching can be causally connected to a merger event that occurred several Gyr earlier, because of the impact that the merger had on the CGM and baryon cycle.

As shown in Figures \ref{fig:merger:bh} and \ref{fig:merger:cgm}, considerable scatter exists in $M_{\rm BH}$ and $f_{\rm CGM}$ between simulations of the same initial conditions with different random number seeds. This scatter propagates into the sSFR of the galaxy and is compounded by the fact that star formation itself is implemented stochastically in these simulations. For the \suppressed{} and \organic{} galaxies, the SFHs of most realisations lie close to the median and indicate present-day star-forming galaxies, though a small number indicate green valley or quenched galaxies at $z=0$. The scatter in the \enhanced{} galaxy's SFH indicates that stochastic effects alone generate a broad range of timescales over which a post-merger galaxy can pass through the green valley and quench; some realisations pass through the green valley within $\approx 700$ Myr of the merger and remain quenched, while others take over 4 Gyr to leave the main sequence and enter the green valley. The distributions shown here demonstrate the importance of accounting for stochastic effects in simulations of individual galaxies. Producing many realisations of each galaxy reveals that the SFHs are distinctly different, but examining only a single realisation of each galaxy could lead to insecure conclusions.

\subsubsection{Stellar kinematics}

\begin{figure}
\includegraphics[width=\columnwidth]{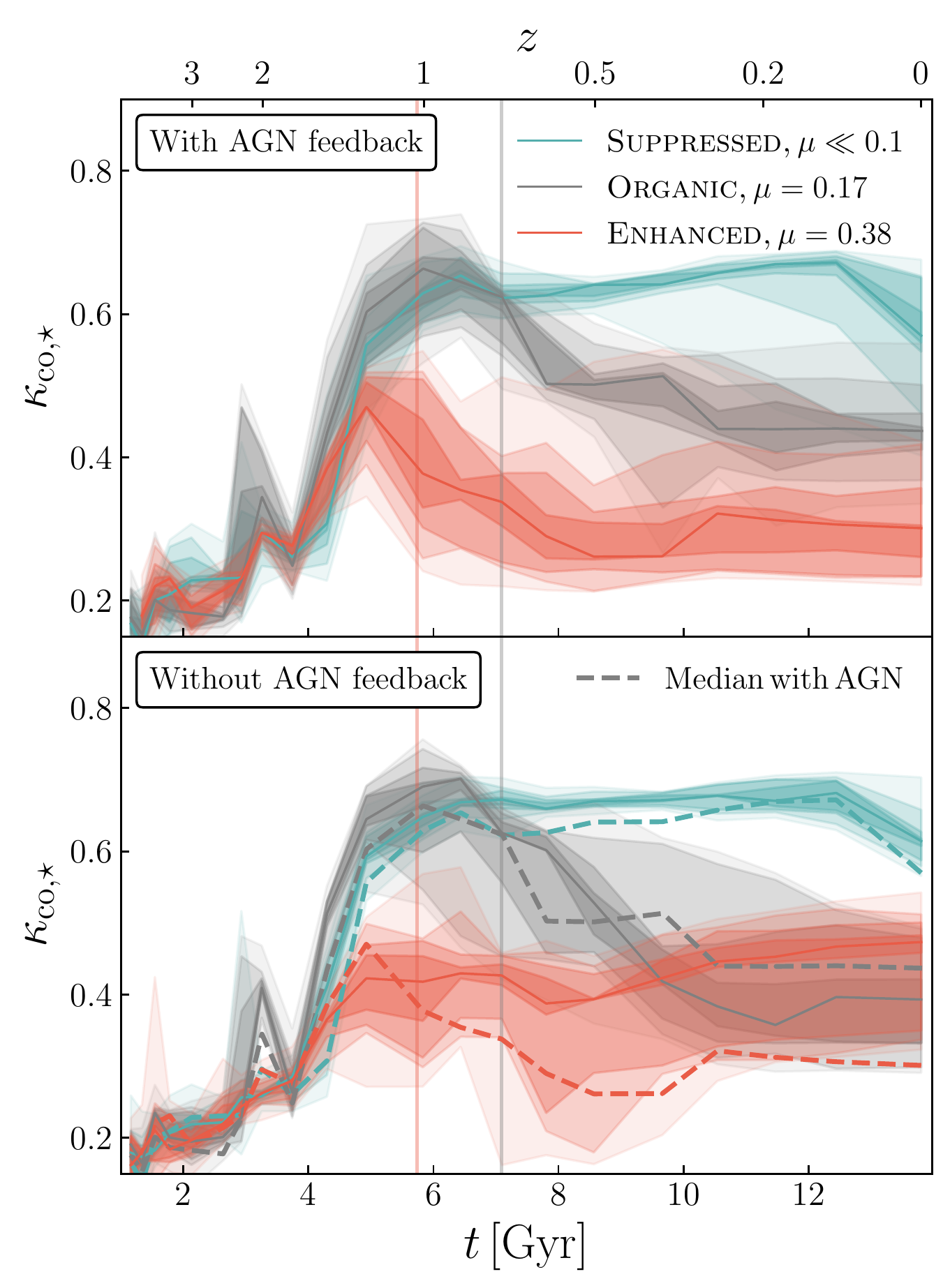}
\caption{Evolution of the stellar co-rotational kinetic energy fraction (\kappacostar{}) for the \organic{} galaxy and its suppressed-merger and enhanced-merger counterparts, shown in the same fashion as Figures \ref{fig:merger:cgm} and \ref{fig:merger:ssfr}. Adjusting the merger ratio also transforms the kinematics and morphology of the disc-like \organic{} galaxy; an enhanced merger yields a dispersion-dominated spheroidal galaxy instead, while a suppresed merger yields a more strongly co-rotating disc. The presence of AGN feedback does not appear to strongly influence the kinematics of the \organic{} and \suppressed{} galaxies, however the \enhanced{} galaxy is able to ``rebuild" its disc to a greater degree when no AGN feedback is present to prevent cooling from the CGM.}
\label{fig:merger:kappa}
\end{figure}

The evolution of \kappacostar{}, shown in the upper panel of Figure \ref{fig:merger:kappa} reveals that adjusting the significance of an individual merger can significantly alter the kinematics of a disc galaxy. At early times, \kappacostar{} increases for all three systems as stars are formed from co-rotating gas. In the \organic{} and \suppressed{} cases the increase in \kappacostar{} slows at $z\sim 1$, when $\kappa_{\rm co,\star}=0.66$ (IQR$=0.13$) and 0.63 (IQR$=0.04$) respectively, both high values characteristic of a star-forming disc galaxy\footnote{A threshold value of $\kappa_{\rm co,\star}=0.4$ has been shown to clearly distinguish star-forming disc galaxies from quenched elliptical galaxies in the EAGLE simulations \citep{correa17}}. The \suppressed{} galaxy experiences no significant mergers capable of disrupting this rotational motion, and \kappacostar{} increases steadily for this system until $t=12.5$ Gyr, after which a mild decline occurs, yielding a present-day $\kappa_{\rm co,\star}=0.57$ (IQR$=0.10$). The minor merger that occurs for the \organic{} galaxy has a disruptive influence on its stellar kinematics, causing a decline in \kappacostar{}, though it remains a co-rotating disc galaxy by $z=0$ ($\kappa_{\rm co,\star}=0.44$, IQR$=0.05$). The head-on major merger experienced by the \enhanced{} galaxy has a far more disruptive effect on its stellar kinematics, with the median \kappacostar{} falling below 0.4 immediately after the merger. The remnant retains kinematics characteristic of an elliptical (or spheroidal) galaxy until the present day, with a final $\kappa_{\rm co,\star}=0.30$ (IQR$=0.12$).

A strong correlation between \kappacostar{} and $f_{\rm CGM}$ has been previously identified within the present-day galaxy populations of cosmological simulations \citep{davies20}, indicating that high-\kappacostar{} disc galaxies are more likely to reside within gas-rich haloes and vice-versa. The results of this study suggest that this correlation could be driven by mergers; galaxies that experience strong mergers are likely to exhibit dispersion-dominated kinematics and spheroidal morphologies for dynamical reasons, and our results here show that mergers can trigger AGN feedback, reducing $f_{\rm CGM}$. However, the extent to which the galaxy kinematics are directly affected by the properties of the CGM remain unclear. 

We can assess the role of the CGM in determining a galaxy's kinematics by comparing the evolution of the \enhanced{} galaxy with and without AGN feedback, as shown in the lower panel of Figure \ref{fig:merger:kappa}. The galaxy is disrupted by a merger in both cases; however, in the absence of AGN feedback the CGM remains gas-rich and able to cool. The galaxy therefore continues to form stars from CGM gas that has coherent angular momentum, thus ``rebuilding" the disc, causing \kappacostar{} to rise and eventually yield a value representative of a disc galaxy. The origin of the $\kappa_{\rm co,\star}-f_{\rm CGM}$ correlations found in cosmological simulations is therefore complex; mergers can disrupt the galaxy and trigger an AGN-driven reduction in $f_{\rm CGM}$, and galaxies in gas-poor haloes are less readily able to gain angular momentum from the CGM. We defer a detailed analysis of the origin of galaxy angular momentum to a future study. 

\section{Summary and Discussion}
\label{sec:summary}

In this study, we have investigated the impact of a galaxy merger on the interaction between active galactic nucleus (AGN) feedback and the circumgalactic medium (CGM) of a simulated Milky Way-like galaxy. We then examined the consequences for the baryon cycle and the evolution of the galaxy. 

To do so, we selected a present-day moderately star-forming (sSFR$=10^{-10.2}$ yr$^{-1}$) central galaxy, with stellar mass $M_\star=4.3\times 10^{10}$ M$_\odot$ and host halo mass $M_{200}=3.4\times 10^{12}$ M$_\odot$ from a simulation of a periodic volume evolved with the Reference EAGLE model \citep{schaye15,crain15} and used zoom initial conditions to re-simulate the galaxy and its environment at 8$\times$ higher mass resolution, adopting EAGLE's recalibrated subgrid parameter values. We then employed the genetic modification (GM) technique \citep{roth16}, using the initial conditions generator \textsc{genetIC} \citep{stopyra20}, to generate two modified variants of our initial conditions, \enhanced{} and \suppressed{}, designed to respectively increase or reduce the disruptive influence of a $z\approx 0.74$ ``target" merger, whilst keeping the overall halo mass history from the fiducial (\organic{}) simulation intact. These modified variants of a single galaxy form a controlled experiment, through which we can establish causal connections between the impact of a galaxy merger and the evolution of the system.

The genetic modifications affect both the stellar mass ratio, $\mu$, of the target merger, and its geometry, quantified by the impact parameter, $r_{\rm impact}$, of the first close interaction. In the \organic{} case, a minor merger occurs, with $\mu=0.17$ and $r_{\rm impact}=0.23r_{200}$, where $r_{200}$ is the halo's virial radius. In the \enhanced{} case, the galaxy instead undergoes a more head-on, major merger, with $\mu=0.38$ and $r_{\rm impact}=0.09r_{200}$. Conversely, in the \suppressed{} case the infalling galaxy enters the primary halo on a tangential orbit, with $r_{\rm impact}=0.75r_{200}$ and does not merge with the central galaxy, remaining a separate object at $z=0$ with $\mu=0.03$. This effectively removes the effect of the target merger. These modifications to the characteristics of an individual merger can be achieved without significantly changing the overall accretion history or final mass of the galaxy's halo (Figure \ref{fig:merger:mass}).

Enhancing the significance of a merger can trigger very rapid growth of the central black hole (BH) and yield a greater final BH mass. Conversely, suppressing a merger reduces the BH accretion rate over long timescales, yielding a lower final BH mass (Figure \ref{fig:merger:bh}). These differing BH accretion histories give rise to significantly different present-day CGM mass fractions; by inducing AGN feedback, a disruptive merger can cause a substantial fraction of the CGM of a Milky Way-like galaxy to be ejected (Figure \ref{fig:merger:cgm}). The rapid BH growth following the \enhanced{} major merger occurs because the rotational motion of gas in the galaxy disc is disrupted, causing gas to collapse towards the centre, elevating the density within the BH kernel. The \organic{} minor merger is not able to disrupt the motion of gas around the BH, and this motion inhibits accretion onto the BH (Figure \ref{fig:vicinity}).

The expulsion of a large fraction of the CGM by AGN feedback in response to a merger causes a significant shift in the density-temperature distribution of the CGM. The gas remaining within the virial radius after a disruptive merger has a lower density and higher temperature than it does when no merger occurs, giving rise to longer cooling times that prevent the galaxy's interstellar medium from being replenished by CGM gas (Figure \ref{fig:phase}). This transformation causes the central galaxy to quench, after a delay of several Gyr corresponding to the timescale for the galaxy's remaining interstellar gas to either form stars, fuel BH growth or be expelled by feedback. Conversely, the more gas-rich CGM of the system that experiences no merger enables a higher sustained sSFR in the central galaxy (Figure \ref{fig:merger:ssfr}). This transformation of the baryon cycle also impacts the central galaxy kinematics; the \enhanced{} galaxy is converted to a dispersion-dominated spheroid by the merger, but it remains a spheroid until $z=0$ because the disc cannot be rebuilt by gas with coherent angular momentum from the CGM (Figure \ref{fig:merger:kappa}).

The significant increase in the BH accretion rate seen in the \enhanced{} galaxy following a major merger is in line with the behaviour of the wider population in the EAGLE simulations. \citet{mcalpine20} found that galaxies hosting AGN are up to 3 times more likely to be undergoing mergers than stellar mass-matched control galaxies without AGN, and that merging galaxies are up to 1.8 times more likely to host an active AGN than non-merging galaxies. We find that the AGN feedback induced by a disruptive merger persists for a very short period of $\approx 0.2$ Myr (Figure \ref{fig:vicinity}); this short timescale offers an explanation for why it is so difficult to associate AGN feedback with mergers in large-volume simulations that are necessarily limited in their output time resolution, and in observations that can only capture an instantaneous snapshot of the galaxy population.

While the intense AGN feedback induced by the merger is very brief, its effects on the evolution of the galaxy-CGM ecosystem unfold over long timescales. Several recent studies have shown that when quenching follows a merger in cosmological simulations it typically does so after a time delay of up to several Gyr \citep[e.g.][]{weinberger18,correa19,rodriguezmontero19,quai21,pathak21}, leading to the notion that mergers and quenching are not causally connected. In our \enhanced{} galaxy, quenching occurs a median $\approx 3$ Gyr after the merger takes place (Figure \ref{fig:merger:ssfr}), yet these events must be causally connected since by construction the parameters of this merger are the primary difference between this quenched galaxy and the star-forming \organic{} galaxy. Our results show that the CGM can mediate this connection over such long timescales; while the feedback induced by the merger expels enough interstellar gas to place the galaxy in the green valley, it is the elevation of the CGM cooling time by this feedback that ultimately prevents further inflow and quenches the galaxy over a longer period. These highly disparate timescales will make it challenging to associate mergers and quenching in both observed and simulated galaxy populations.

Observational indicators of past morphological transformation such as the spin parameter ($\lambda$) and the ratio of ordered to disordered stellar velocity ($V/\sigma$) have been found to be poor predictors of quenching relative to the overall stellar velocity dispersion \citep{brownson22}. While we find that the kinematics of our \enhanced{} galaxy are transformed by a merger, this takes several Gyr to complete (Figure \ref{fig:merger:kappa}) and it is not clear that such a transformation is necessary for the galaxy to quench. The key action of the merger is to introduce a short-term {\it local} change to the motion of gas in the BH vicinity (Figure \ref{fig:vicinity}), and such brief disruptions may be sufficient to unlock rapid BH growth in galaxies without changing the {\it global} kinematics of the stars and gas.

This study was performed within the context of the EAGLE simulation model, and it is important to consider how our results might differ if another model were used. The most critical difference is likely to be in how efficiently AGN feedback can impact the CGM as a function of the BH accretion rate, $\dot{m}_{\rm accr}$, and the BH mass, $M_{\rm BH}$. The EAGLE model implements AGN feedback with a single mode in which energy injection rate into the CGM scales with $\dot{m}_{\rm accr}$. In a contrasting example, the two-mode AGN feedback employed by the IllustrisTNG model can only expel gas from the CGM in a ``kinetic mode'', which is active when the Eddington ratio is low and the BH mass is greater than $\approx 10^{8.2}$ M$_\odot$ \citep[][]{terrazas20}. Despite these differences, the two models can yield similar correlations between the properties of galaxies and those of their gaseous environments \citep[e.g.][]{davies20} because the key quantity for the CGM is the total AGN feedback energy injected in a mode capable of expelling gas. In EAGLE, the high accretion rate induced by the \enhanced{} merger drives efficient feedback, but also increases the BH mass relative to the \organic{} case, facilitating higher rates still (since $\dot{m}_{\rm accr}\propto M_{\rm BH}^2$). It is possible that in a two-mode model akin to that in IllustrisTNG, the action of this rapid accretion is to increase the BH mass until efficient kinetic feedback commences, thus achieving a similar effect on the CGM but through a different route. A direct code comparison would be required to determine whether this is the case.

With our controlled study, we have shown for the first time how disruptive events such as galaxy mergers are capable of driving strong changes in the content and properties of the CGM through the AGN feedback they induce. A key question remains: are such events {\it required} for AGN feedback to significantly impact the CGM of Milky Way-like central galaxies, driving diversity in their properties? This study shows that the BH mass, CGM mass fraction and sSFR are sensitive to individual merger events; however, these properties are also strongly correlated with the intrinsic properties of the host dark matter halo \citep{davies19,davies20,davies21}. Haloes that assemble earlier at a given mass tend to host quenched, CGM-poor galaxies because they are quicker to reach the threshold mass at which BHs can grow at a non-linear rate \citep{bower17,mcalpine18}, and because they are more concentrated and confine gas at the halo centre more effectively, yielding a more massive BH and the injection of more AGN feedback \citep{boothschaye10,boothschaye11}. 

There are two possible scenarios in which merger events and the intrinsic properties of the dark matter halo can both influence the evolution of the galaxy-CGM ecosystem:
\begin{enumerate}
    \item Merger events {\it are} required for AGN feedback to transform the baryon cycle and quench galaxies, but the quiescent galaxies residing in early-assembling haloes have almost always experienced such a merger.
    \item Merger events {\it are not} required, and early halo assembly alone is sufficient for enough BH growth and AGN feedback to transform the baryon cycle and quench galaxies. Galaxies residing in later-assembling haloes are typically star-forming at the present day, but can end up quenched if they experience a disruptive merger and undergo the process seen in our \enhanced{} system.
\end{enumerate}

Establishing which of these scenarios is correct will be challenging. Galaxies residing in haloes with differing assembly histories will experience different merger histories as a natural consequence of hierarchical assembly, and many degenerate factors determine the impact of a merger on a galaxy, such as the geometry \citep[e.g.][]{zeng21}, mass ratio, gas richness and redshift. Controlling for these variables makes it difficult to assemble samples of galaxies that have (or have not) experienced {\it disruptive} mergers in statistical populations, or to construct a GM experiment in which the halo assembly time is varied for a fixed merger history.

Taken in isolation, the results of this study appear to favour scenario (i). Our haloes have near-identical assembly times and exceed the critical mass for non-linear BH growth at $z\approx 2$, yet the \suppressed{} system, which evolves secularly after this time, experiences suppressed BH growth and very little CGM ejection. A disruptive merger appears to be required for the baryon cycle to be transformed in this system. Our results reveal the importance of mergers in inducing BH growth in an EAGLE galaxy which already hosts a massive BH; \citet{mcalpine18} also note that at low redshift, mergers can be important in initiating non-linear growth of EAGLE BHs at or near the seed mass. Further evidence supporting scenario (i) exists in the present-day EAGLE and IllustrisTNG populations; early-assembling, gas poor haloes typically host spheroidal galaxies with low \kappacostar{} \citep{davies20}, suggesting that these systems may have experienced a disruptive event that unlocked the growth the the BH in the past\footnote{We note that galaxies residing in earlier-assembling haloes could also exhibit low \kappacostar{} simply because their haloes have less time to gain angular momentum from gravitational torques before turnaround \citep[e.g.][]{zavala16}, and not because they are more likely to experience disruptive mergers.}. These correlations are relatively mild, however, and there are many undisturbed systems in gas-poor haloes, lending some credence to scenario (ii). 

While the relative importance of mergers and disruptive events in driving diversity in the properties of the wider galaxy population remains unclear, we have shown in this study that they can profoundly affect the galaxy-CGM ecosystem when they occur, and that their importance may be underestimated due to the timescales involved. The small-scale, short-timescale processes of gas disruption and BH growth that they induce can drive a large-scale transformation of the baryon cycle, the consequences of which can take several Gyr to unfold and culminate in quenching.

\section*{Acknowledgements}
We thank the referee for their constructive suggestions that improved this manuscript. JJD thanks Ted Mackereth, Corentin Cadiou, Stephen Stopyra and Gandhali Joshi for helpful discussions that greatly contributed to this work. This study was supported by the European Union's Horizon 2020 research and innovation programme under grant agreement No. 818085 GMGalaxies. AP and RAC are supported by the Royal Society. This study used computing equipment funded by the Research Capital Investment Fund (RCIF) provided by UKRI, and partially funded by the UCL Cosmoparticle Initiative. It also made use of high performance computing facilities at Liverpool John Moores University, funded by the Royal Society and LJMU’s Faculty of Engineering and Technology. It utilised the publicly-available EAGLE galaxy catalogues \citep{mcalpine16}, and analysis was performed using \textsc{pynbody} \citep{pontzen13} and \textsc{tangos} \citep{pontzentremmel18}.

\section*{Data Availability}

The data underlying this article will be shared on reasonable request to the corresponding author.



\bibliographystyle{mnras}
\bibliography{bibliography}



\bsp	
\label{lastpage}
\end{document}